\documentclass{article}

\usepackage{arxiv}

\usepackage[utf8]{inputenc} 
\usepackage[T1]{fontenc}    
\usepackage{hyperref}       
\usepackage{url}            
\usepackage{booktabs}       
\usepackage{amsmath,amsfonts,amssymb}       
\usepackage{nicefrac}       
\usepackage{microtype}      
\usepackage{graphicx}
\usepackage[backend=biber, style=numeric, sorting=none, citestyle=numeric]{biblatex}
\usepackage{graphicx}%
\usepackage{multirow}%
\usepackage{tabularx}
\usepackage{array,multirow,graphicx}
\usepackage{makecell}
\usepackage{doi}

\title{Meta-optical Imaging at Thermal Wavelengths}

\author{Anna Wirth-Singh$^{1,*}$\\
\And
Aurelia M. Brook$^{1}$
\And
Rose Johnson$^{2}$
\And
Johannes E. Fr\"{o}ch$^{1,2}$\\
\And 
Arka Majumdar$^{1,2,**}$\\
\And
\\
$^{1}${Department of Physics, University of Washington, Seattle, WA, USA}\\
$^{2}${Department of Electrical and Computer Engineering, University of Washington, Seattle, WA, USA}\\
$^{*}$\textit{annaw77@uw.edu}\\
$^{**}$\textit{arka@uw.edu}
}

\begin{document}
\maketitle

\begin{abstract}
The field of meta-optics has grown to include metalenses spanning the ultraviolet to terahertz regimes. Imaging is a key application of meta-optics, with recent works demonstrating meta-optical imaging with advanced functionalities including wide field of view, broadband operation, and polarization sensitivity. In this review, we focus on meta-optical imaging for thermal wavelengths. Thermal meta-optics are less well-studied than those in the visible range but have vast potential applications spanning defense, health, and geological sensing. We first introduce these applications and their specific challenges. With compact form-factor and multi-functional capabilities, we suggest that meta-optics are particularly well-suited to meet the needs of imaging in the mid- and long-wave infrared. Then, we review published experimental demonstrations of thermal imaging via meta-optics. These meta-optics vary in complexity from simple hyperboloid metalenses to complex systems composed of engineered meta-atoms and multiple layers of optics. Finally, we suggest some areas where thermal meta-optics may be useful, and we identify some emerging approaches to solve lingering challenges of meta-optical imaging. 
\end{abstract}

\section{\label{sec:Introduction}Introduction}
Enabled by advancements in optics and sensor technologies, it is possible to image the world at wavelengths spanning almost the whole electromagnetic spectrum. Each wavelength range has particular advantages for particular applications. For example, X-ray imaging is useful for biomedical applications, but is seldom applied in daily life, while cell phone cameras operating in the visible are ubiquitous to capturing scenes in daily life, but are limited to daytime or well-lit situations. Towards longer wavelengths, the mid-wave infrared (MWIR, 3 to 5 $\mu$m) and long-wave infrared (LWIR, 8 to 14 $\mu$m) ranges are often termed ``thermal” wavelengths. According to Planck's law of blackbody radiation, objects emit a spectrum of light according to their temperature, with hotter objects emitting at greater intensity and with a spectral distribution shifted towards shorter wavelengths. Objects at common temperatures have high contrast with the ambient background in the MWIR to LWIR, and this unique property makes the thermal ranges indispensable for applications spanning consumer electronics to defense, including night vision \cite{Waxman98}, non-contact thermography \cite{Brzezinski21}, and long-range atmospheric, geological, and agricultural sensing \cite{Ishimwe14,Wilson2023,Havens16}.

In most, if not all, thermal imaging applications, the size and weight of the imaging system may be highly constrained. For example, night vision goggles require compact form-factor and light weight to avoid causing fatigue and discomfort for the user \cite{Parush11}. Excess weight is also unacceptable in many long-range sensing applications, where the imaging system is mounted on an airborne unit or drone. Many defense applications require discreet and compact imaging systems that can be integrated into larger, multi-functional devices. However, high-quality imaging typically requires compound lens assemblies, so there is a trade-off between form-factor and optical performance. Especially for thermal wavelengths, these complex assemblies are expensive and difficult to manufacture due to the limited availability of suitable refractive lens materials. Therefore, improving the form-factor and practical applicability of thermal imaging systems is an active area of research. 

\begin{figure}[h!]
\centering\includegraphics[width=15cm]{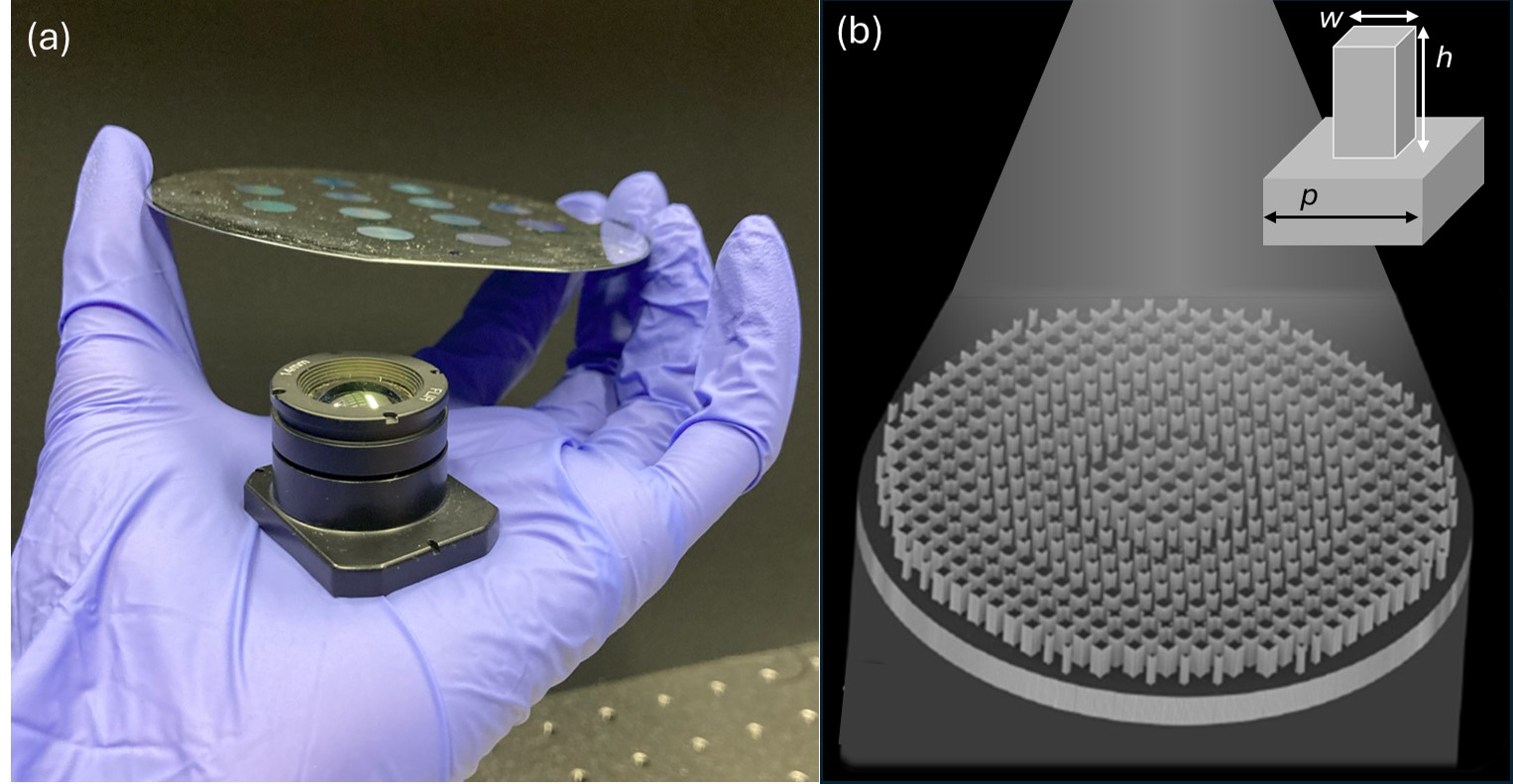}
\caption{Operating principles of meta-optics. (a) A photograph comparing a typical LWIR refractive lens assembly to meta-optics. The pictured silicon wafer contains several fabricated meta-optics. (b) Schematic illustration of a meta-optic. The inset highlights one scatterer unit cell. The unit cell consists of a pillar of width $w$ and height $h$ arranged in a lattice of periodicity $p$. The height and periodicity are typically fixed while the width or cross-sectional shape is varied to adjust the phase delay. \textit{Reproduced from \cite{Huang24} under \href{https://creativecommons.org/licenses/by/4.0/}{CC.BY 4.0}}}
\label{Fig:Overview}
\end{figure} 

Interest in ultra-thin and flat meta-optics has surged in recent years due to their potential to miniaturize and enhance existing imaging systems. Meta-optics are quasi-periodic arrays of sub-wavelength scatterers, or pillars, that impart a local phase shift to the light. By engineering the arrangement of scatterers across the surface, a global phase distribution can be realized to perform functions including, but not limited to, imaging\cite{Tseng21,Chen20,Colburn18fullcolor,Shrestha18}, sensing \cite{Froech22}, and optical computing \cite{Colburn19,Zheng22,WirthSingh24onn}. Including the substrate upon which the pillars are supported, meta-optics are typically only 200 $\mu$m to 500 $\mu$m in thickness. Figure \ref{Fig:Overview}a shows a 300 $\mu$m thick silicon wafer containing several LWIR meta-optics and a compound refractive lens assembly to highlight the thinness of meta-optics compared to the refractive counterparts. Many inherent challenges in meta-optical imaging, such as strong chromatic aberrations \cite{Wang17} and limited field of view (FoV) \cite{Mart20}, now have solutions. Motivated by consumer-facing applications in compact cameras \cite{Tseng21,Pinilla23} and augmented/virtual reality \cite{Li21,Li22Inverse,Lee18,Bayati21}, significant work has been undertaken in the field of meta-optics for imaging at visible wavelengths in particular. The principles guiding design and operation of thermal meta-optics are the same as those that apply to visible meta-optics, with adjustments to physical size scaling and material platforms. Investigation into thermal meta-optics is especially warranted because meta-optics are extremely well-suited to the unique imaging applications of this wavelength range. Compared to the visible imaging range, applications in the thermal range are more defense-focused, emphasizing compact form-factor and multi-functionality over high-quality full-color imaging. Therefore, we argue that meta-optical imaging is naturally aligned with the needs of thermal imaging applications.

In this report, we review the field of meta-optics for thermal imaging. We first introduce the thermal wavelength ranges, their imaging applications, and the specific challenges associated with them. This includes a brief overview of meta-optics and their operating principles, but note that several other review articles \cite{Yu14,Chen16,Huang22,Chen20} provide a more detailed review on the operating principles of meta-optics. Other forms of flat optics, including multilevel diffractive optics, have been demonstrated at thermal wavelengths \cite{Meem19,Hayward23mdl}; however, we limit this discussion to sub-wavelength meta-optics specifically. Then, we review the state-of-the art in meta-optics for thermal imaging, broadly organized into three categories: hyperboloid imaging metalenses, meta-optics for imaging at wide field of view, and meta-optics for broadband imaging. Table \ref{Tab:Summary} summarizes these works. To conclude, we suggest where future thermal meta-optics efforts may be focused to maximize their impact in real-world applications.  

\begin{table}
\caption{\label{Tab:Summary}Reported experimental demonstrations of thermal imaging with meta-optics. }
\footnotesize
\begin{tabular}{@{}lllllll}
\bottomrule
Ref. & Design Merits  & Wavelength & Material &  Aperture & Focal Length & Full FoV \\
\midrule
\multicolumn{7}{l}{\textbf{Hyperboloid Meta-Optics}} \\
\\

\cite{Zuo17} & High Efficiency & 4.0 $\mu$m & $\alpha$-Si:H on MgF$_2$ & 300 $\mu$m & 50 - 300 $\mu$m & 53$^{\circ}$ $^{*}$ \\
\cite{Fan18} & High NA & 10.6 $\mu$m & All-Si & 12 mm & 8 mm & 74$^{\circ}$ $^{*}$ \\
\cite{Huang21} & Ambient Illumination & 10.0 $\mu$m & All-Si & 2 cm & 2 cm & 53$^{\circ}$ $^{*}$ \\
\cite{Nalbant22} & High Efficiency & 9.07 $\mu$m & All-Si with ZnS & 2 cm & 2 cm & 53$^{\circ}$ $^{*}$ \\
\cite{Li22} & Large Aperture  & 10.0 $\mu$m & All-Si & 80 mm & 80 mm & 37$^{\circ}$ $^{*}$ \\
\cite{Saragadam24} & Foveated Imaging & 10.0 $\mu$m & All-Si doublet & \makecell[l]{75 mm \\ 25 mm} & \makecell[l]{150 mm \\ 25 mm} & 53$^{\circ}$ $^{*}$ \\
\midrule

\multicolumn{7}{l}{\textbf{Wide FoV Meta-Optics}} \\
\\
\cite{Shalaginov20} & External Aperture & 5.2 $\mu$m & PbTe on  CaF$_2$ & \makecell[l]{1 mm aperture \\ 5.2 mm optic} & 2 mm & $180^\circ$ \\
\cite{WirthSingh23} & External Aperture  & 10.0 $\mu$m & All-Si & \makecell[l]{1 cm aperture \\ 3 cm optic} & 1 cm & $80^\circ$ \\
\cite{Zhao23} &  Aperture Array & 10.0 $\mu$m & All-Si & \makecell[l]{1.2 mm \\(each, 5 total)} & 2.4 mm & $60^\circ$ \\
\cite{Lin24} & External Aperture & 10.6 $\mu$m & All-Si & \makecell[l]{1 cm aperture \\ 4.8 cm optic} & 2 cm & $140^\circ$ \\
\cite{WirthSingh2024zoom} & Zoom Imaging & 3.4 $\mu$m & c-Si on Al$_2$O$_3$ & 8 mm & 8.8 - 44 mm & $50^\circ$ \\
\midrule

\multicolumn{7}{l}{\textbf{Broadband Meta-Optics}} \\
\\

\cite{Ou2021broadband} & Dispersion Engineering  & 3.5 - 5 $\mu$m & All-Si & \makecell[l]{200 $\mu$m \\ 370 $\mu$m} & \makecell[l]{400 $\mu$m \\ 200 $\mu$m} & 53$^{\circ}$ $^{*}$ \\
\cite{Huang24} & Inverse Design & 8 - 12 $\mu$m & All-Si & 1 cm & 1 cm & 53$^{\circ}$ $^{*}$ \\
\cite{Liu24hybrid} & Hybrid Refractive & 8 - 12 $\mu$m & All-Si & 6.3 mm & 13 mm & 20$^{\circ}$  \\

\bottomrule
\end{tabular}\\
$^{*}$FoV was not provided, but was estimated from provided numerical aperture (NA). 

\end{table}
\normalsize

\section{Thermal Imaging Applications}

\begin{figure}[h!]
\centering\includegraphics[width=15cm]{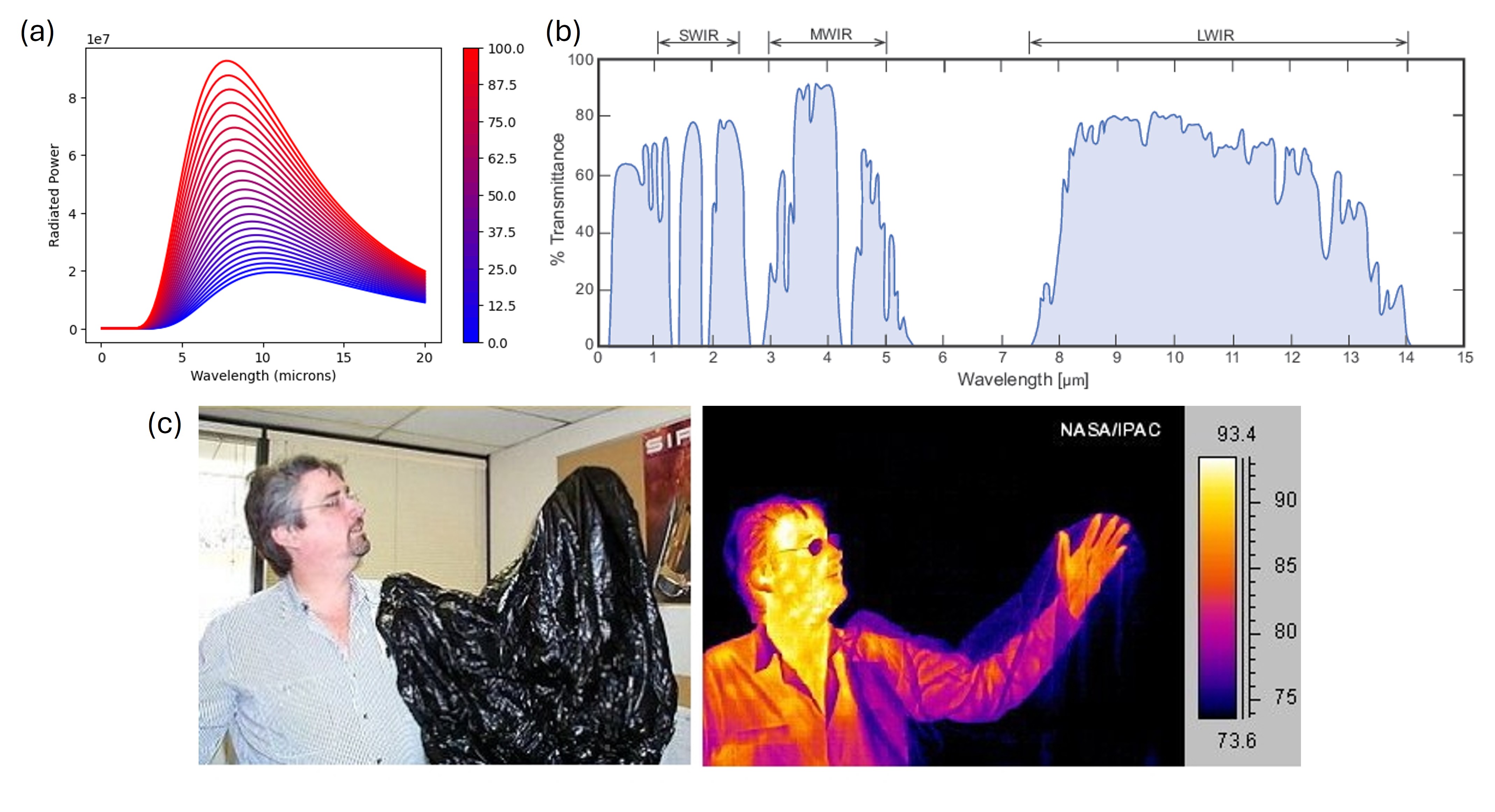}
\caption{Imaging at thermal wavelengths. (a) Blackbody radiation spectrum as a function of temperature, from 0 to 100$^\circ$C. (b) Atmospheric transmission spectrum. \textit{Reproduced from \cite{RadarSystems2013}.} (c) A scene imaged using a visible camera (left) versus a thermal infrared camera (right). A person's hand is behind a thin sheet of plastic, which is opaque at visible wavelengths but transparent under thermal wavelengths. This illustrates the utility of thermal imaging to view objects that are obscured at visible wavelengths but visible at thermal wavelengths.}
\label{Fig:Blackbody}
\end{figure}

According to Planck’s Law, a thermal blackbody at temperature $T$ emits electromagnetic radiation according to the spectrum

\begin{equation}
    B(\lambda,T) = \frac{2\pi h c^2}{\lambda^5} \frac{1}{e^{\frac{h c}{\lambda k_B T}}-1}
\end{equation}

where $\lambda$ is the emitted wavelength, $c$ is the speed of light, $h$ is Planck's constant, and $k_B$ is Boltzmann's constant. This spectrum is shown for various temperatures in Figure \ref{Fig:Blackbody}a. Objects at common ambient temperatures of 0 - 100$^\circ$C emit in the LWIR range, with the spectrum shifting towards shorter wavelengths for hotter objects. For realistic objects, the intensity of radiation is also related to the emissivity $\epsilon$, which is a property of the material. A perfect blackbody with $\epsilon = 1$ emits the maximum amount of energy at a given temperature. A highly reflective metallic material, such as aluminum, has $\epsilon \approx 0.2$. 

The atomospheric transmission spectrum is shown in Figure \ref{Fig:Blackbody}b. Between MWIR (3 - 5 $\mu$m) and LWIR (8 - 14 $\mu$m), from around 5 to 8 $\mu$m, water in the atmosphere absorbs radiation. Applications within this intermediate range are therefore limited. While there are many similarities between MWIR and LWIR imaging, there are advantages and disadvantages for each range that influence their utility for particular applications. Only very hot objects emit below 3 $\mu$m, which is why the lower range of thermal imaging is typically considered to be around 3 microns. While LWIR imaging relies primarily on emission from terrestrial objects at ambient temperatures, MWIR imaging leverages some reflection from hotter sources, such as sunlight and starlight, on objects in addition to their own thermal emission \cite{Dhar08}. For this reason, LWIR is more suitable for thermometry applications at typical ambient temperatures. On the other hand, MWIR imaging typically offers higher resolution due to the shorter wavelength and operating principle of the sensor. MWIR wavelengths penetrate high humidity weather conditions more effectively than LWIR, making MWIR particularly attractive for long-range sensing \cite{Dhar08}. However, MWIR cameras tend to be more expensive and the signal to noise ratio for cooler objects is better in the LWIR range, so the best choice ultimately depends on the application. 

Thermal imaging was initially used for military applications \cite{Wilson2023} and is still heavily utilized in the aerospace and defense industry. Specifically, it is used for navigation and targeting, on drones, and weapon scopes, and especially for night vision. Night vision is typically enabled by one of two methods. In the near-infrared band (750 - 1550 nm), reflected moon and star light is amplified via intensifier tubes and converted to visible illumination \cite{Waxman98}. Due to the many elements required - NIR optics, an intensifier, and VIS optics - night vision systems employing this method are typically bulky. Alternatively, night vision based on thermal emission requires no additional illumination for compact and robust night imaging systems. In particular, thermal-based night vision allows for clearer imaging in situations where light pollution and weather would obscure NIR-based viewing \cite{Estrera03}.

In recent times, non-contact thermography enabled by LWIR cameras was widely used during the Covid-19 pandemic to detect fever as a symptom of transmittable illness \cite{Brzezinski21,Martinez20,Cardwell20}. In most use-cases, a hand-held device is used to image a person and convert their thermal emission to a temperature reading. Such applications benefit from compact and lightweight cameras. In one proposed application \cite{Barnawi21}, an aerially-mounted thermal camera captures infrared images of a crowd, which are subsequently postprocessed to identify individuals who may be ill with Covid-19. Similar schemes have been proposed for ecological monitoring and health assessments, where remote sensing systems are minimally disruptive to wildlife \cite{Lavers05,Havens16}. Aerially-based remote sensing is also used in agriculture to identify plant health and water sources. In these applications, long-range imaging at high resolution is required while maintaining low weight for drone-mounted imaging.

While advancements in sensor technologies have made thermal imaging accessible to local governments and commercial entities, thermal imaging sensors are still significantly more expensive than visible sensors. Typically, LWIR cameras are microbolometers which convert heat to electricity. In this way, LWIR cameras may be compact and do not require external cooling, but suffer from limited resolution due to heat diffusion throughout the sensor. MWIR cameras, on the other hand, typically employ a cooling system. While on-board cooling provides higher resolution for better imaging performance, it also contributes to the significant cost of MWIR camera sensors. Since thermal sensors are expensive, it is practical to design multi-functional optics rather than purchase additional sensors for different applications. Furthermore, due to the limited resolution of the sensor itself, the requirements of thermal imaging optics are typically less stringent on resolution and place greater emphasis on functions such as magnification factor and field of view. These are both areas in which meta-optics can excel.

\section{Thermal Optics Materials and Fabrication}

Imaging lenses are required to be transparent at the wavelength range of interest and simultaneously provide sufficient index contrast with the air. Transparent materials at thermal wavelengths are generally limited to silicon, germanium, and chalcogenide glasses (such as CaF$_2$, ZnSe, and MgF$_2$). Single crystalline germanium is often preferred due to its favorable chromatic and thermal properties, but the material is rare and expensive \cite{Zhang03}. A concern unique to thermal imaging applications is that the lens itself may become heated, for example in high-temperature environments or under laser illumination, leading to increased thermal emission that adds haze to the image \cite{Desnijder23}. This may be a significant issue for bulk refractive optics, but is less problematic for meta-optics due to their thin form-factor that quickly dissipates excess heat.

Several material platforms have been developed for meta-optics at visible wavelengths, including thin crystalline silicon \cite{Li19}, silicon nitride \cite{Zhan16}, and TiO$_2$ \cite{Khorasaninejad17} on glass or fused silica substrates. However, different materials are required for thermal meta-optics due to the poor transmission of these materials in the thermal wavelength range. All-silicon is a popular platform for thermal meta-optics, particularly at LWIR wavelengths. In contrast to silicon-on-sapphire or silicon-on-glass substrate, wherein the contrasting substrate provides a hard stop for etching, the continuous silicon layer allows for etching to any height by performing a timed etch; this provides maximum design flexibility and avoids the challenges of depositing a thick film on substrate. As an example of this platform's versatility, there are several reports of all-silicon meta-optics for imaging around 10 $\mu$m wavelength with differing pillar geometries. Fan et al. \cite{Fan18} use 6.8 $\mu$m tall circular pillars situated on a lattice with 6.2 $\mu$m periodicity, while Huang et al. \cite{Huang21} use 10 $\mu$m tall square-shaped silicon pillars situated on a lattice with 4 $\mu$m periodicity and Li et al. \cite{Li22} use 7 $\mu$m tall square-shaped pillars on a lattice with 5.8 $\mu$m periodicity. While the aspect ratio of pillars is practically limited due to structural stability, a report in the near-infrared \cite{Lim21} demonstrates the possibility of using hole-based optics to increase the possible aspect ratio for greater control over chromatic dispersion. Hole-based optics have also been proposed for LWIR applications \cite{Nalbant22}.

While the all-silicon platform has many benefits, including large aspect ratios, CMOS-compatible fabrication, readily available materials, and low cost, the high refractive index contrast between the substrate and the air results in high Fresnel reflection loss. Applying an anti-reflection coating is an effective way to reduce this loss, with zinc sulfide (ZnS) being the preferred choice at thermal wavelengths \cite{Cox58}. Recently, it has been shown that such a single layer of ZnS coating increases the efficiency of a germanium-based meta-optic from 75\% to 97\% \cite{Nalbant22}. Alternatively, a thin layer of silicon on a transparent dielectric substrate such as BaF$_2$ \cite{Li22} or Al$_2$O$_3$ \cite{WirthSingh2024zoom} can improve transmission as compared to an all-silicon substrate. 

In order to fabricate sub-wavelength structures, high-resolution lithography tools are required. However, the longer wavelengths of MWIR and LWIR relax fabrication constraints as compared to visible meta-optics. Visible meta-optics are almost exclusively patterned using electron beam lithography, which is capable of producing features as small as a few tens of nanometers but is also slow and expensive. More scalable ultraviolet light-based methods, including direct write lithography and deep ultraviolet stepper lithography, have also been used to fabricate meta-optics \cite{Park24glass,Kim23DUV} and other nanostructured devices \cite{Tao21DUV}. While the minimum resolution of some DUV facilities (around 250 nm) is limiting for visible meta-optics, this resolution is sufficient for thermal meta-optics. Therefore, thermal meta-optics are amenable towards mass production and large aperture optics.

\section{Hyperboloid Metalenses}

To function analogously to a spherical lens and provide diffraction-limited focusing, a flat lens must have a hyperboloid phase profile given by \cite{Aieta12}:

\begin{equation}
    \phi(r) = -\frac{2\pi}{\lambda}(\sqrt{r^2+f^2}-f)
\end{equation}

where $r$ is the radial coordinate, $f$ is the desired focal length, and $\lambda$ is the design wavelength in free space. A flat lens with a hyperboloid phase profile provides diffraction-limited focusing at normal incidence and single-wavelength illumination; however, performance deteriorates quickly away from these ideal conditions. Hence, a different design strategy is required for wide FoV or broadband operation, as discussed in the later sections. 

The hyperboloid phase profile has historically been used as baseline for first demonstrations and as a point of comparison for new materials and novel design strategies. In particular, several early demonstrations explored new material platforms supporting high efficiency. In the MWIR, Zuo et al. \cite{Zuo17} fabricated and characterized hyperboloid meta-optics based on hydrogenated amorphous silicon ($\alpha$-Si:H, n = 3.5) pillars on MgF$_2$ (n = 1.37) substrate for MWIR applications. With this platform, the high refractive index contrast between the pillars and the substrate supports strong field confinement in the pillars for high focusing efficiency, and the relatively low substrate index minimizes reflections for high transmission. The authors fabricated and characterized a set of metalenses in this platform with constant focal length of 300 $\mu$m and varying aperture sizes corresponding to NAs from 0.45 to 0.95. In each case, the  size of the measured spot is in good agreement with the theory (nearly diffraction-limited) and the imaging performance was similar as compared to a traditional molded aspheric chalcogenide lens \cite{Zuo17}. In the fabricated metalenses, an experimentally measured focusing efficiency of $\approx78\%$ was reported. While the relatively high-index pillars and low-index substrate of this work supports higher efficiency, the more complex material stack leads to greater cost. 


\begin{figure}[h!]
\centering\includegraphics[width=15cm]{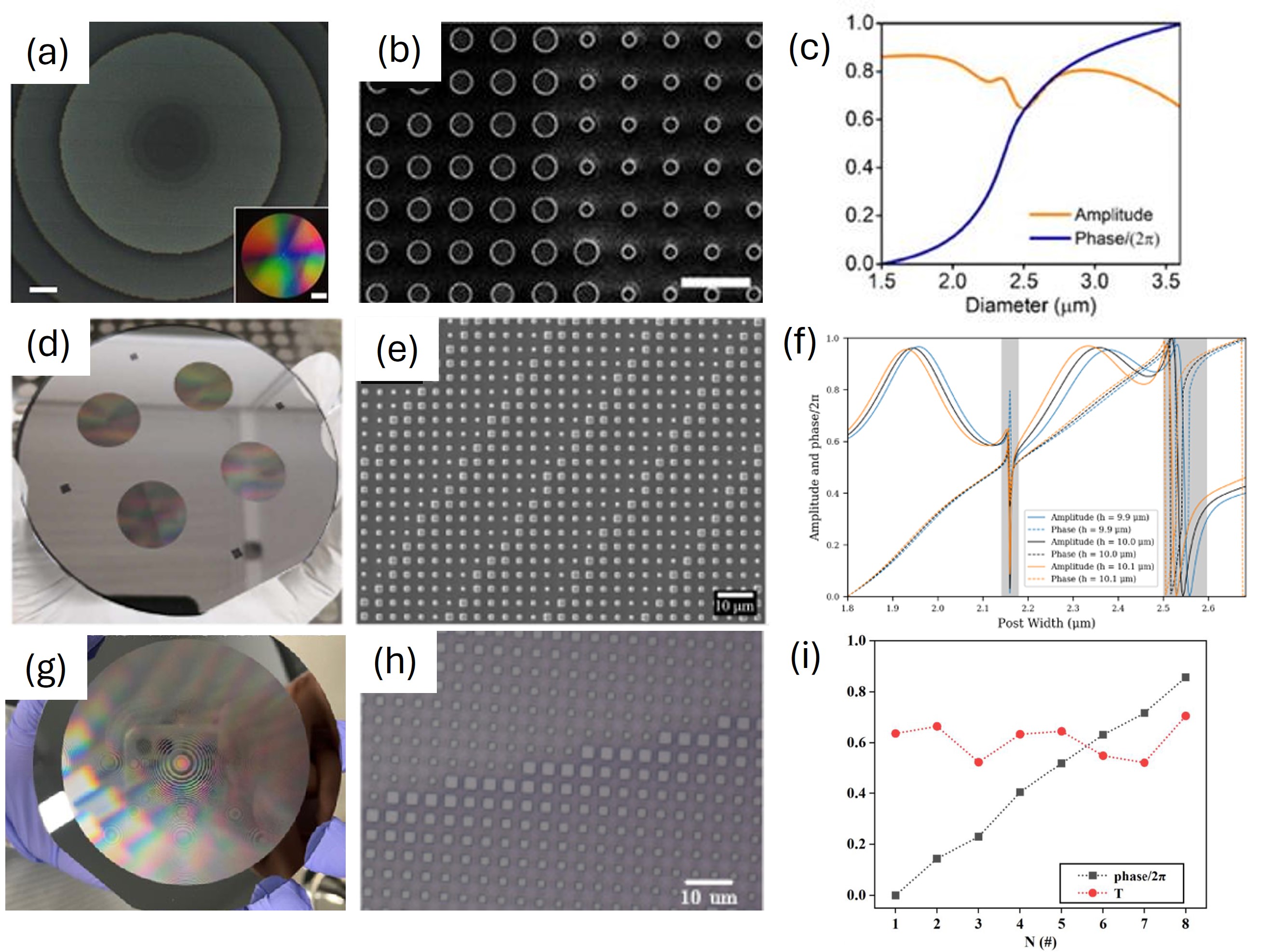}
\caption{LWIR Hyperboloid meta-optics. (a) Optical microscope image of a 12 mm diameter meta-optic. The scale bar is 100 $\mu$m. The inset shows a photograph of the entire fabricated optic with scale bar of $2$ mm. (b) SEM image of the fabricated meta-optics. The scale bar is 10 $\mu$m. (c) Simulated amplitude and phase for the pillar library with $p = 6.2$ $\mu$m and $h = 6.8 $ $\mu$m. (a - c) \textit{Reproduced from \cite{Fan18}}. (d) Photograph of 2 cm aperture meta-optics on a 100 mm diameter wafer. (e) SEM image of the fabricated meta-optics with scale bar of 10 $\mu$m. (f) Simulated amplitude and phase for the pillar library with $p = 4.0$ $\mu$m and $h = 10.0$ $\mu$m. (d - f) \textit{Reproduced from \cite{Huang21}}. (g) Photograph of 80 mm diameter meta-optic. (h) SEM image with scale bar 10 $\mu$m. (i) Simulated phase and transmission for the pillar library with $p = 5.8$ $\mu$m and $h = 7$ $\mu$m. (g - i) \textit{Reproduced from \cite{Li22}}. }
\label{Fig:HypLWIR}
\end{figure}

In LWIR, early work focused on hyperboloid phase profile lenses in all-silicon material platforms, which is advantageous as a readily available material and is compatible with CMOS fabrication processes. Fan et al. \cite{Fan18} demonstrated a 12 mm aperture, NA = 0.6 metalens at 10.6 $\mu$m operating wavelength. Under laser illumination, nearly diffraction-limited focusing was reported. Similarly, Huang et al. \cite{Huang21} demonstrated a 2 cm aperture, NA = 0.45 metalens at 10.0 $\mu$m operating wavelength. These optics are shown in Figure \ref{Fig:HypLWIR}. In contrast to the laser-illuminated imaging demonstrations in Fan et al., Huang et al. demonstrate imaging under ambient thermal radiation conditions. Despite blurring due to the strongly chromatic nature of hyperboloid metalenses, this work demonstrated that ambient light imaging is possible with these simple metalenses. 

\begin{figure}[h!]
\centering\includegraphics[width=15cm]{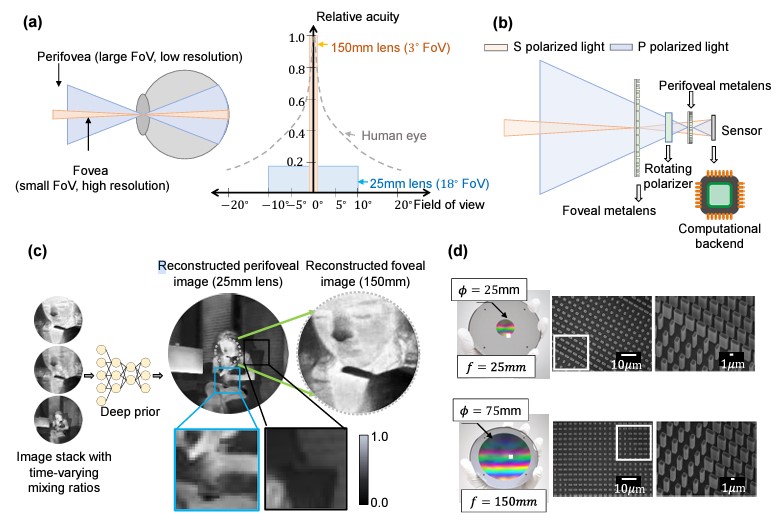}
\caption{Foveated thermal imaging system using meta-optics. (a) The foveated metalens is inspired by the human eye, which has high acuity (high resolution) over a small FoV (fovea) and a low acuity over a large FoV (perifoveal). (b) The optic consists of a foveal element at a focal length of 150 mm, a perifoveal element at 25 mm, and a freely rotating polarizer. (c) A simulated example illustrating the image reconstruction algorithm. (d) Photographs (left) and scanning electron microscope images (right) of the fabricated meta-optics. \textit{Reproduced from \cite{Saragadam24}. }}
\label{Fig:foveated}
\end{figure}

More recent work has focused on the development of large aperture metalenses. Large aperture optics are often desirable to improve the signal-to-noise ratio and facilitate imaging over long distances, but large aperture refractive lenses are bulky, heavy, and difficult to manufacture. Due to the thinness of meta-optics, the aperture can be increased without incurring substantial weight, making this an attractive application area for thermal meta-optics. Recently, Li et al. \cite{Li22} demonstrated an ultra large aperture (80 mm diameter) hyperboloid metalens for LWIR imaging under ambient illumination for a nominal wavelength of 10 $\mu$m. Similar to earlier LWIR metalenses, this metalens was fabricated using direct-write lithography in all-silicon platform, consisting of square pillars of height 7 $\mu$m and lattice periodicity 5.8 $\mu$m. This slightly larger periodicity accommodates fabrication constraints inherent in fabricating such a large area metalens, and the single-step lithography and CMOS compatible processing enable a path towards low-cost, large aperture thermal lenses based on meta-optics.

Another recent work by Saragadam et al. \cite{Saragadam24}, summarized in Figure \ref{Fig:foveated}, highlights an example of a multi-functional meta-optics to achieve foveated imaging. Inspired by the human eye, foveated imaging combines a high-resolution but narrow FoV center image for detail and a low-resolution, wide FoV image for context. This concept is particularly applicable to thermal wavelengths, where limited detector pixel size makes simultaneous capture of high resolution and wide FoV especially challenging. To demonstrate foveated imaging, Saragadam et al. design a three-element optical system consisting of two polarization-sensitive hyperboloid metalenses and a rotating polarizer. To achieve polarization sensitivity, they use silicon pillars with rectangular cross-section. Each metalens images only one polarization to a designed focal length (and hence FoV), and the dynamic mixing ratio between the two polarization is controlled by the polarizer. Finally, a computational reconstruction algorithm based on a deep generative prior is used to reconstruct the images.

\section{Meta-optics with Wide Field of View}

Surveillance and imaging applications often benefit from a wide field of view, which allows a wide scene to be captured in a single camera frame without rotating the camera. Generally, ``wide" angle lenses are considered to be those with a FoV greater than 60$^{\circ}$ and ``ultrawide" or ``fisheye" lenses are those with a FoV greater than around 100$^{\circ}$ \cite{Laikin80,Kumler00}. To achieve this FoV using refractive optics, multiple lenses are stacked to correct for Seidel aberrations that worsen at increasing field angle. This stacking of multiple lenses contributes to the size and weight of the optical system, which is undesirable for size- and weight-constrained applications. 

The aforementioned hyperboloid metalens phase profile is unsuitable for imaging at wide FoV; even for incident angles of only $2^{\circ}$, distortion is evident in hyperboloid metalenses \cite{Mart20}. However, with other design approaches, meta-optics are well-suited to wide FoV imaging and an overview of these approaches and the fundamental principles is reviewed in Ref. \cite{Yang23}. One way to achieve wide FoV using meta-optics is to use a quadratic phase profile \cite{Pu17}. A quadratic phase profile mimics a spherical lens in the limit of infinite radius and infinite refractive index to provide good (but no longer diffraction-limited) resolution at wide FoV \cite{Mart20}. In many applications, especially those at thermal wavelengths, achieving diffraction-limited resolution may not be required and wide FoV would be a preferred attribute. The quadratic phase profile approach has been used to demonstrate meta-optics with an impressive 170$^{\circ}$ FoV at visible wavelengths \cite{Mart20} and could be readily adapted to thermal imaging applications. 

\begin{figure}[h!]
\centering\includegraphics[width=13cm]{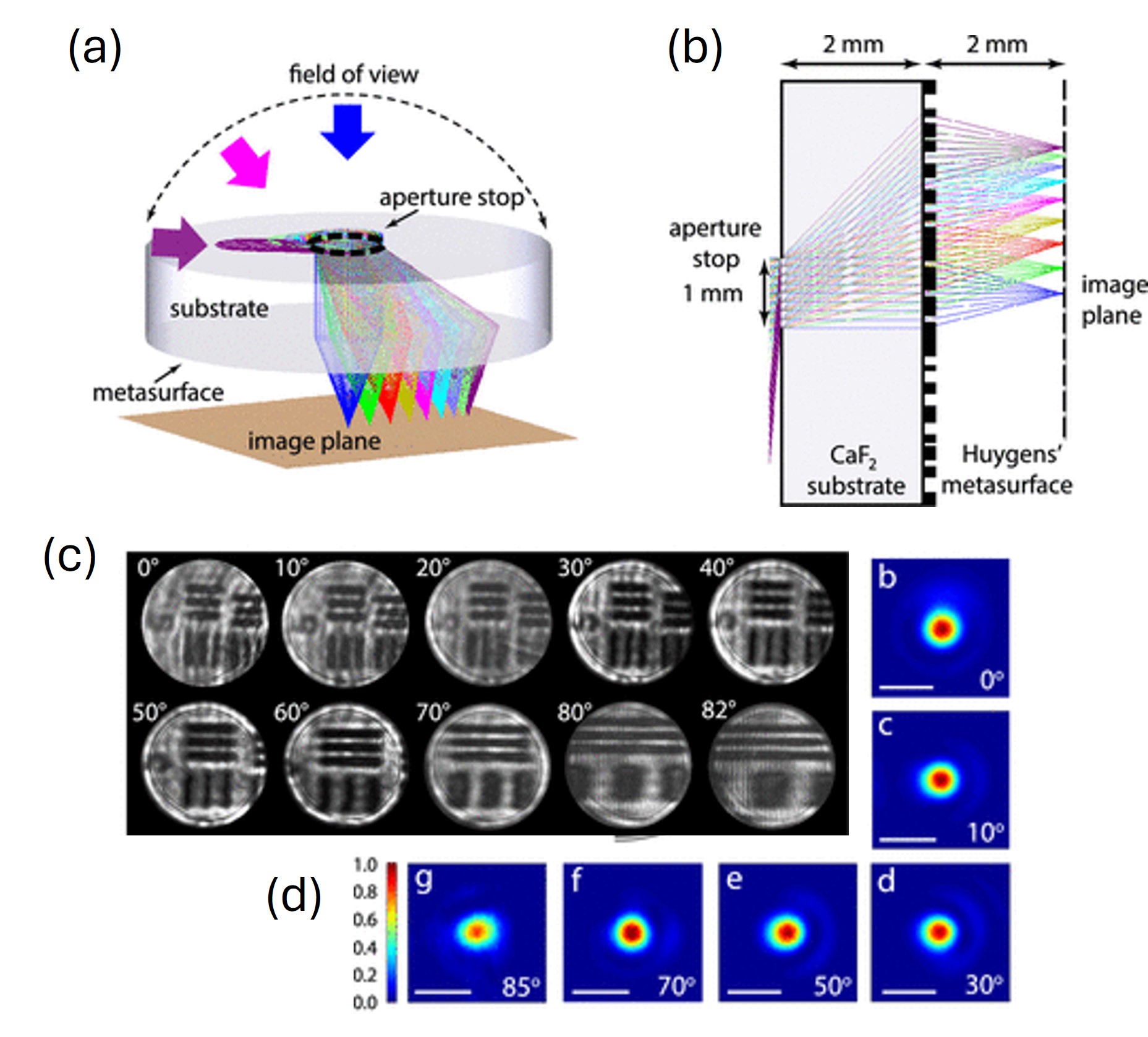}
\caption{Wide FoV meta-optic for imaging in the MWIR. (a) Three-dimensional schematic illustrating the aperture concept. (b) A two-dimensional ray tracing diagram of the wide FoV MWIR meta-optic. (c) Experimental imaging results. The results show the same groups from the USAF resolution chart, under laser illumination, for angles of incidence up to $82^\circ$. (d) Measured point spread functions up to $85^\circ$ angle of incidence. \textit{Reproduced with permission from \cite{Shalaginov20}}.
}
\label{Fig:large_FoV_mwir}
\end{figure}

Another approach towards achieving wide FoV is to use an external aperture to restrict incident light such that light of different incident angles interacts with different sections of the lens. Therefore, each part of the lens can be designed specifically for a particular angle of incidence. This idea is schematically depicted in Figure \ref{Fig:large_FoV_mwir}a, and wide FoV meta-optics combined with an external aperture have been demonstrated at MWIR \cite{Shalaginov20} and LWIR \cite{WirthSingh23,Lin24} wavelengths. When using this approach, the phase profile of the meta-optic is typically optimized using a commercial ray tracing software (e.g., Zemax OpticStudio) that models the meta-optic as a radially symmetric phase mask parameterized by coefficients that are optimized to minimize spot size or wavefront error over large angles of incidence. If the aperture is sufficiently small (and therefore, the beams of different incoming angles are sufficiently separated), then diffraction-limited resolution can be achieved. This approach has been demonstrated at MWIR wavelengths \cite{Shalaginov20}, for impressive diffraction-limited imaging up to 170$^{\circ}$. In this work, summarized in Figure \ref{Fig:large_FoV_mwir}, the entrance aperture is 1 mm ($\approx$200$\lambda$) and the optic itself is 5.2 mm ($\approx$1000$\lambda$) in diameter. Here, the optic is comprised of Huygens meta-atoms \cite{Zhang18} made of PbTe resting on a CaF$_2$ substrate, but the concept for achieving wide FoV by using the external aperture is scalable to other meta-atoms and wavelength bands.

\begin{figure}[h!]
\centering\includegraphics[width=15cm]{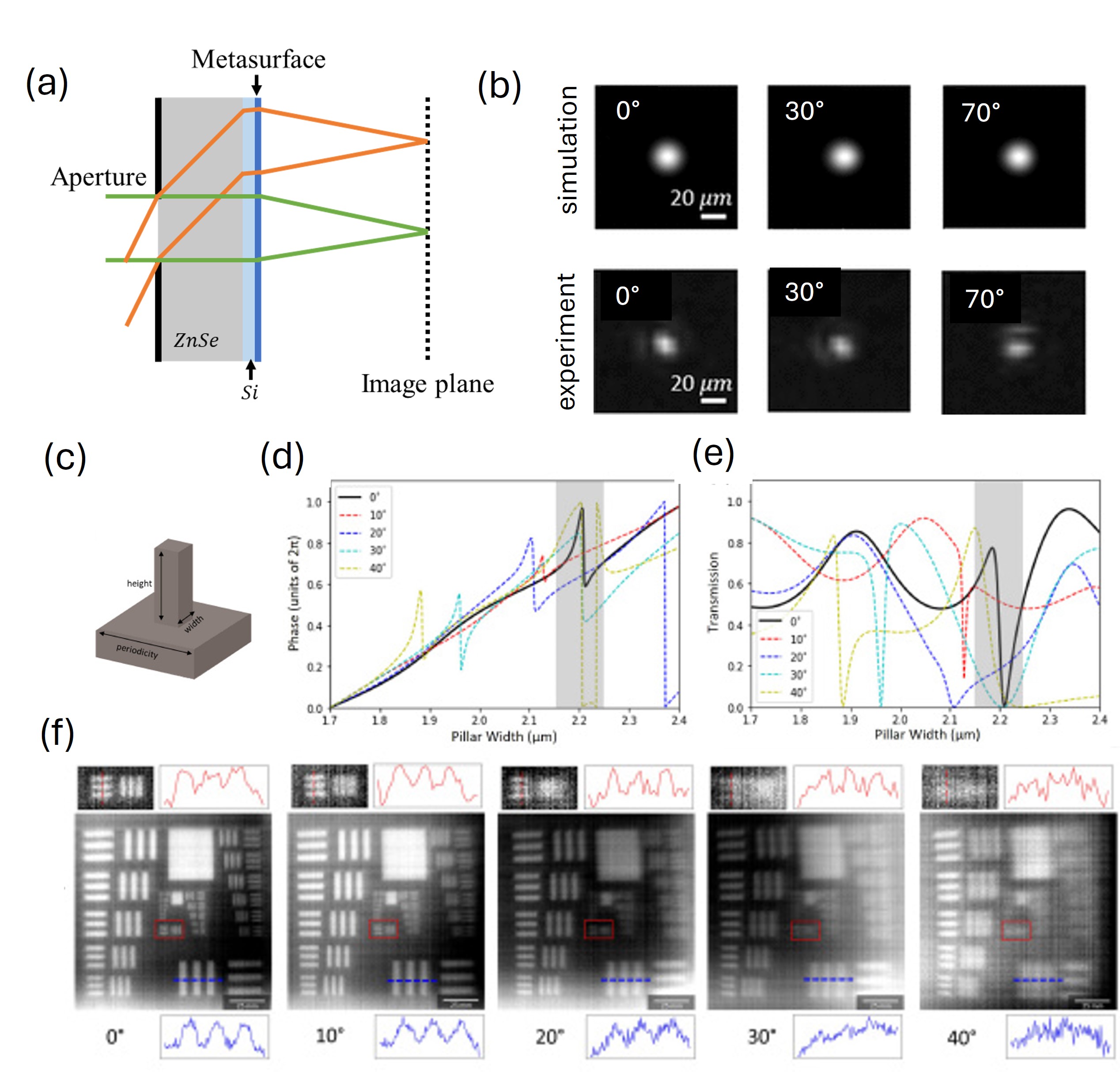}
\caption{Wide FoV meta-optics for imaging in the LWIR. (a) Schematic of the concept utilizing an external aperture and ZnSe spacer between the aperture and metasurface. (b) Simulated and experimental PSFs of a wide FoV LWIR metalens. (a - b) \textit{Reproduced from \cite{Lin24}}. (c) Scatterer unit cell diagram. (d) Simulated phase and transmission (e) of the unit cell under angles of incidence from $0^\circ$ to $40^\circ$. (f) Experimental imaging results at increasing angles of incidence. (c-f) \textit{Reproduced with permission from \cite{WirthSingh23}}.
}
\label{Fig:large-FoV-aperture}
\end{figure}

However, there is an inherent trade-off in the aperture-based approach to achieving wide FoV. While reducing the entrance aperture enables greater spatial separation of the light for higher-resolution imaging, a reduced aperture also limits the amount of light entering the optical system for reduced signal-to-noise ratio. Especially for ambient imaging in low-light or long-range conditions, which is often the case in thermal imaging, a very small entrance aperture may be unacceptable. With this trade-off in mind, two recent works have demonstrated wide FoV imaging at LWIR wavelengths using an entrance aperture around 1 cm ($\approx$1000$\lambda$); these reports are summarized in Figure \ref{Fig:large-FoV-aperture}. Wirth-Singh et al. \cite{WirthSingh23} demonstrate imaging up to 80$^{\circ}$ full FoV that exceeds the performance of a comparable refractive singlet lens at large angles of incidence. While the optic was designed for single-wavelength imaging at 10 $\mu$m, imaging under narrowband ($10 \pm 0.25$ $\mu$m) and broadband ($8-14$ $\mu$m) illumination is also demonstrated, with the broadband result being improved with simple computational postprocessing techniques. In another demonstration, Lin et. al \cite{Lin24} demonstrate imaging up to 140$^{\circ}$ using a similar approach involving an external aperture. In addition, this work employs a ZnSe (n = 2.4) spacer between the aperture and meta-optic to enhance performance over a comparable design utilizing an air gap between the aperture and the meta-optic. Specifically, the high-index spacer reduces the rapid divergence of rays at large angles of incidence, which is a key limitation on attainable FoV. While the addition of a high-index spacer enables wider FoV, it should also be kept in mind that a non-air spacing element increases the weight of the optical system.

\begin{figure}[h!]
\centering\includegraphics[width=12cm]{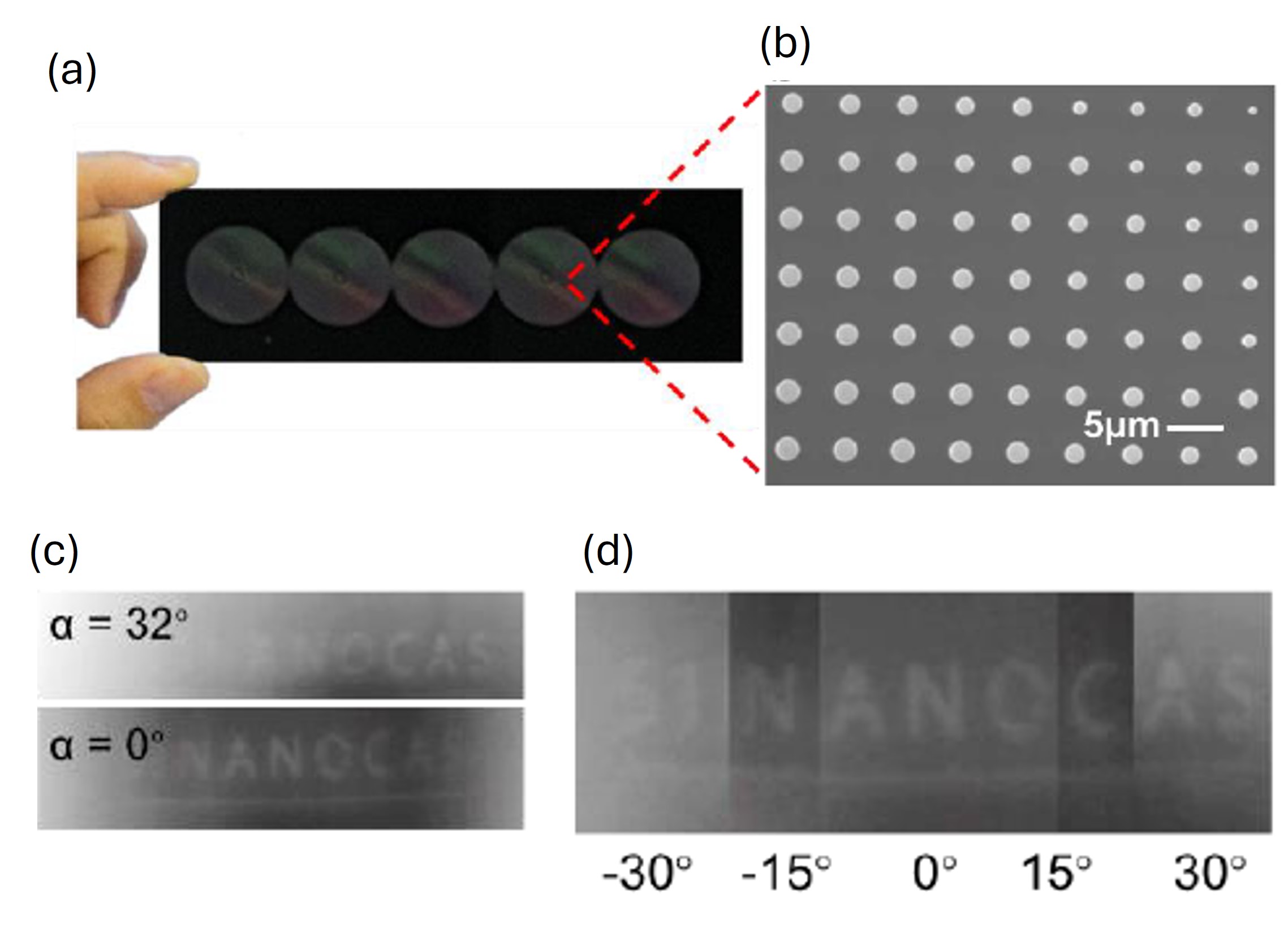}
\caption{Wide FoV LWIR imaging via an aperture array. (a) A photograph of the fabricated $5\times 1$ array of meta-optics. (b) Scanning electron microscope image of the fabricated optics. (c) Exemplary images from $0^\circ$ and $32^\circ$ angle of incidence. (d) The wide FoV imaging result, composed of the images from the corresponding sub-apertures stitched together. \textit{ Reproduced from \cite{Zhao23}}.
}
\label{Fig:lwir_array}
\end{figure}

A restricted aperture is a common characteristic of wide FoV imaging systems because it is required to conserve etendue \cite{Yang23}.  While the single quadratic phase metalens \cite{Mart20} may appear to circumvent this, a virtual aperture exists in these systems \cite{Yang23}. Hence, the use of an external aperture is a common approach. In this implementation, however, the gap between the aperture and the meta-optic necessarily increases the thickness of the optical system and does not fully utilize the inherent thinness of meta-optics. An alternative to the external aperture approach for wide FoV is a metalens array. In a metalens array, several metalenses are arranged on a single substrate, each with its own aperture. Therefore, the restricting aperture has been moved into the same plane as the meta-optics for a truly flat wide FoV system, but some computational techniques (e.g. stitching) are required to reconstruct the image. This approach has been demonstrated in the visible \cite{Chen22aperture}, in the NIR \cite{Hu24}, and also in the LWIR \cite{Zhao23}. In the LWIR, Zhao et al. \cite{Zhao23} derive an expression for the required phase of each sub-lens and use a 1D array of 5 sub-lenses to accomplish $60^\circ$ full FoV imaging. In this case, each sub-lens is a modest 1.2 cm in diameter and has 2.4 cm focal length. Imaging covering the full FoV is demonstrated by stitching together the imaging results from the sub-lenses, as shown in Figure \ref{Fig:lwir_array}.

To maintain high resolution with a large aperture and wide FoV, one possibility is to correct aberrations using additional layers of meta-optics. In a doublet configuration, the second layer of meta-optics serves as both an aperture and a corrective plate to improve performance as compared to an aperture alone. This has been accomplished at near-infrared \cite{Arba16} and visible \cite{Groever17,WirthSingh24eyepiece} wavelengths, and could be adapted for thermal applications as well. Further, one recent work by Wirth-Singh et al. \cite{WirthSingh2024zoom} reports a triplet system to accomplish zoom imaging with up to $50^\circ$ full FoV in the mid-infrared. By axially translating two of three meta-optics in the system, the magnification of the system is adjusted for up to $5\times$ zoom factor. These works demonstrate the ability of meta-optics to replace traditional refractive optics in increasingly complex, multi-layer wide FoV optical systems.

\section{Meta-optics for Broadband Thermal Imaging}

Strong chromatic behavior is a characteristic of meta-optics and other forms of diffractive optics \cite{Huang22, Presutti20}. This effect can greatly deteriorate imaging performance, as it is impossible to focus light at different wavelengths in the same focal plane. A number of works in the visible have addressed this issue using dispersion engineering \cite{Chen20,Shrestha18}, computational imaging \cite{Bayati21EDOF,Tseng21}, and hybrid meta-optic/refractive systems \cite{Pinilla23}. While chromatic aberrations affect thermal meta-optics as well, these issues may be more easily mitigated in thermal imaging settings. Contrary to visible cameras, thermal imaging sensors are typically monochromatic. Therefore, chromatic aberrations manifest as a haze absent of color information. In this way, the chromatic issues may be easier to eliminate using simple computational postprocessing techniques. Furthermore, several thermal meta-optics reports have demonstrated that reasonable broadband imaging results can be achieved without explicitly designing for it. Huang et al. \cite{Huang21} demonstrated that, despite the chromatic nature of the hyperboloid metalens profile, imaging under ambient illumination is possible. Further, Wirth-Singh et al. \cite{WirthSingh23} use simple de-noising algorithms \cite{Maki20} to improve the quality of broadband imaging in a system designed at a single wavelength. In particular, they show that broadband illumination with computation achieves qualitatively comparable results to narrowband illumination using a filter. 

\begin{figure}[h!]
\centering\includegraphics[width=15cm]{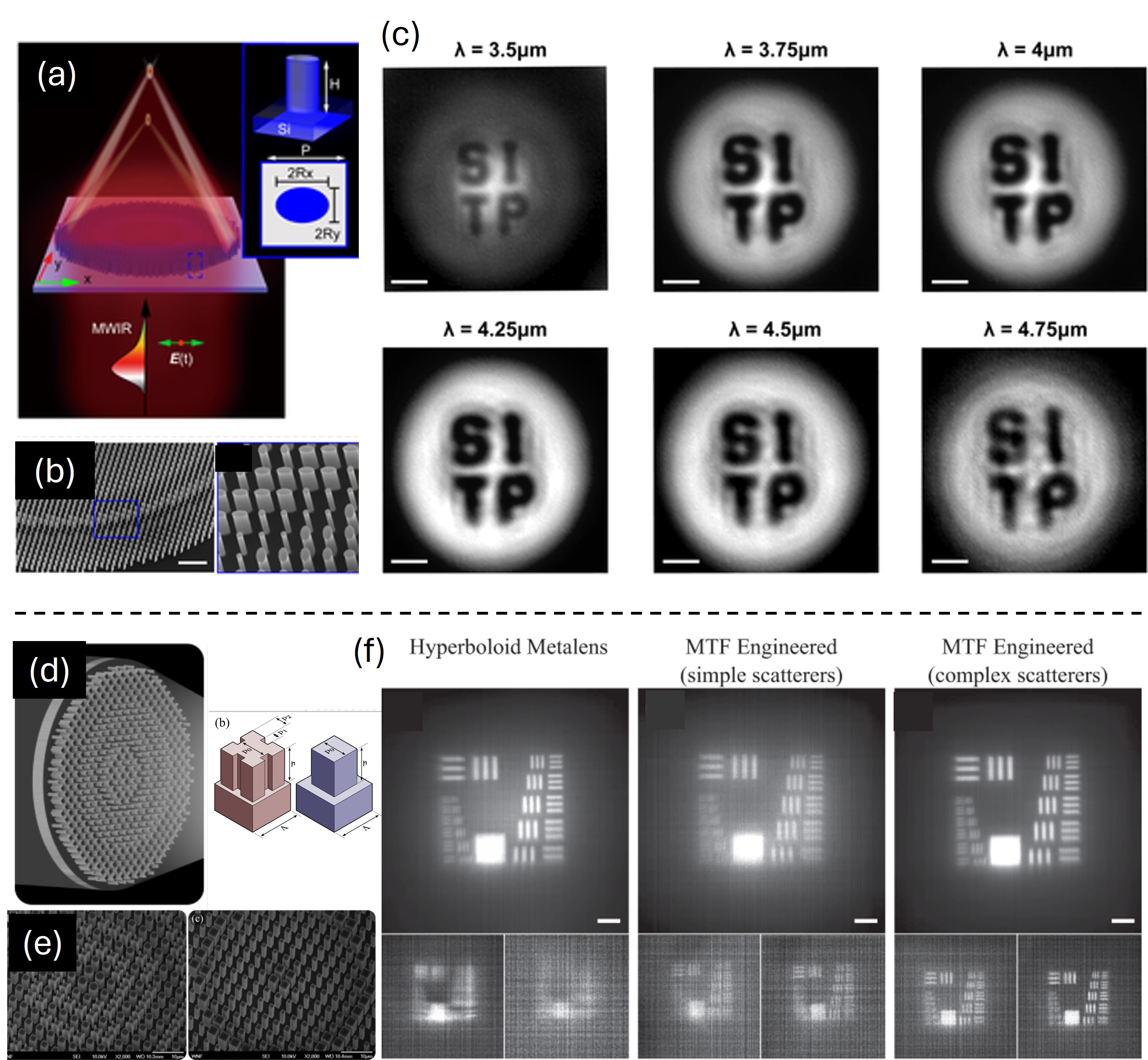}
\caption{Broadband imaging results. (a) Schematic of varifocal MWIR broadband imaging using dispersion engineering. The inset shows a schematic of a meta-atom unit cell.  (b) Scanning electron microscope images of fabricated optics consisting of elliptical-shaped pillars. (c) Experimental imaging results at a range of wavelengths. (a-c) \textit{ Reproduced from \cite{Ou2021broadband}}. (d) Schematic of LWIR broadband imaging using simple and complex-shaped meta-atoms using MTF engineering. (e) Scanning electron microscope images of the complex (left) and simple (right) meta-optics. (f) Experimental broadband imaging results using a hyperboloid meta-lens (left), MTF-engineered meta-optic with simple scatterers (center), and MTF-engineered meta-optic with complex scatterers (right). Underneath each larger image is the result with the addition of bandpass filters at $10 \pm 0.25$ $\mu$m and $12 \pm 0.25$ $\mu$m. (d-f) \textit{Reproduced with permission from \cite{Huang24}}.
}
\label{Fig:broadband-1}
\end{figure}

By explicitly designing for broadband operation, better broadband imaging quality can be achieved. The dispersion engineering approach for broadband meta-optics employs complex-shaped scatterers to increase phase diversity of the scatterer library. Fabricating such complex structures can be challenging in the visible range, requiring resolution on the scale of tens of nanometers. At thermal wavelengths, where the wavelength is an order of magnitude larger, similar dispersion engineering approaches can be accomplished with 100 nm features. However, in order to achieve sufficient dispersion for broad bandwidth operation, very large aspect ratio pillars are often required. With the dispersion engineering approach, there are fundamental trade-offs between bandwidth, NA, and device thickness \cite{Presutti20}; this generally limits dispersion-engineered meta-optics to small NA and small apertures. A number of works have reported design and simulation of broadband MWIR \cite{Yue23,Zhou19bband,Li21bband,Guo22bband,Shih22} and LWIR \cite{Shan22bband,Zhao23bband,Yue23} meta-optics using dispersion engineering, but few reports include experimental verification. Ou et al. \cite{Ou2021broadband} experimentally demonstrate both polarization-insensitve and polarization-sensitive broadband metalenses over 3.5 $\mu$m to 5 $\mu$m bandwidth in all-Si material platform. For polarization sensitivity, their design utilizes elliptically shaped meta-atoms and they use this sensitivity to demonstrate varifocal imaging, wherein the x- and y-polarizations provide different focal lengths. Via point spread function measurements, they demonstrate less than 6\% deviation in the focal length relative to the mean and 45\% focusing efficiency \cite{Ou2021broadband}.
 
In a different approach, Huang et al. \cite{Huang24} developed a multi-scale inverse design framework to simultaneously optimize meta-atom geometry and the global phase profile for broadband imaging in the LWIR. In an inverse design process for large area meta-optics, it is not computationally feasible to simulate the interaction between the light and the pillars at each iteration. To solve this problem, they simulated a library of meta-atoms and used the simulated data to train a deep neural network. The deep neural network enables a quick mapping between meta-atom and phase within the optimization loop. They apply this framework to design broadband LWIR meta-optics with 1 cm aperture, both with simple square pillars and complex cross-shape meta-atoms in an all-silicon platform. Under broadband LWIR illumination, the imaging quality of meta-optics using complex pillars is improved; these results are summarized in Figure \ref{Fig:broadband-1}. Using this design framework, they report a factor of six improvement in wavelength-averaged Strehl ratio as compared to a standard hyperboloid metalens \cite{Huang24}.

\section{Outlook}

\begin{figure}[h!]
\centering\includegraphics[width=15cm]{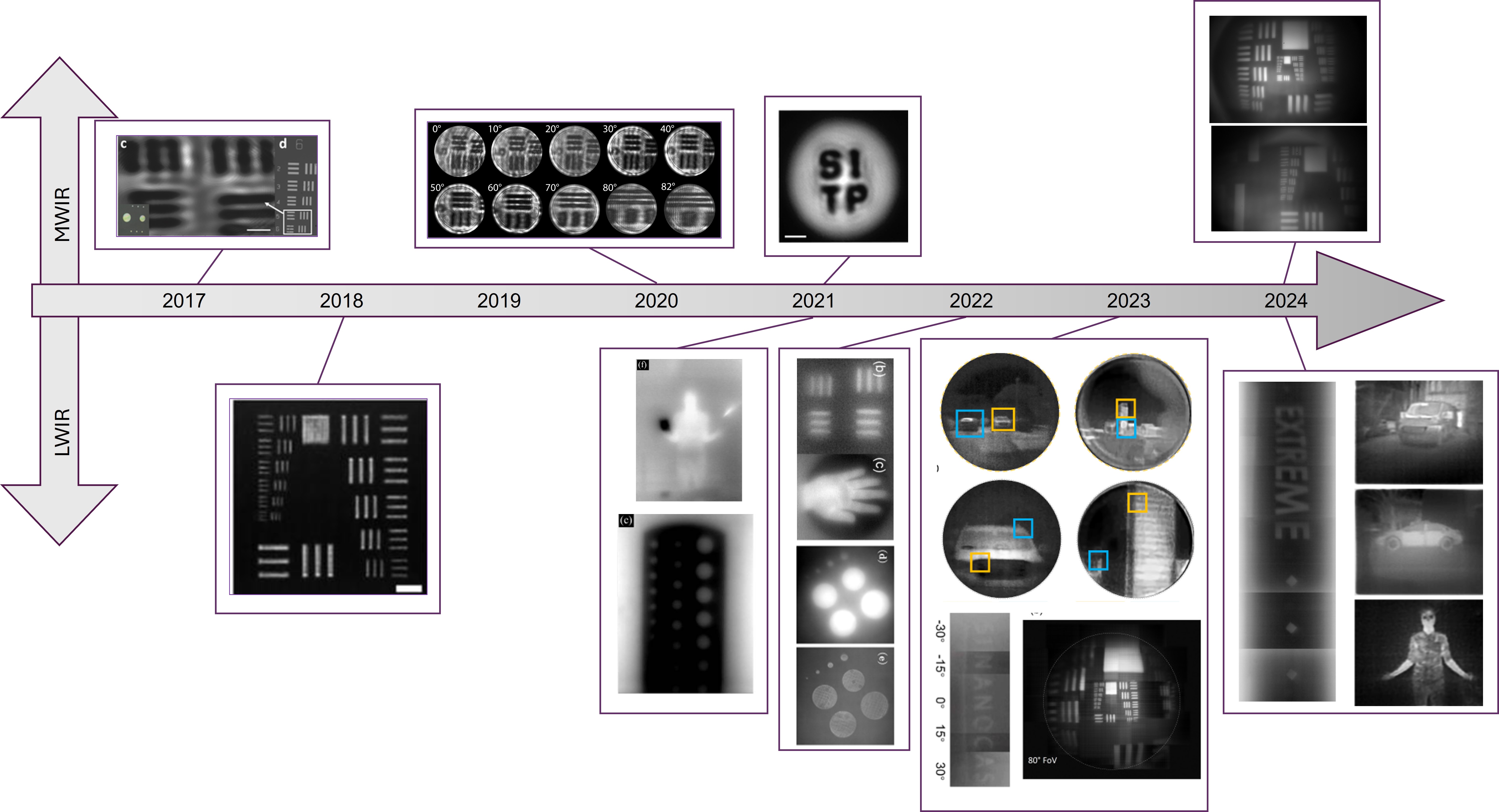}
\caption{Progression of thermal imaging using meta-optics. Demonstrations in the MWIR are shown above, and demonstrations in the LWIR are below. Images reproduced with permission from: \cite{Zuo17} (2017), \cite{Fan18} (2018), \cite{Shalaginov20} (2020), \cite{Huang21,Ou2021broadband} (2021), \cite{Nalbant22,Li22} (2022), \cite{Saragadam24,Zhao23,WirthSingh23} (2023), and \cite{WirthSingh2024zoom,Huang24,Lin24}. }
\label{fig_progression}
\end{figure}

In Figure \ref{fig_progression}, we compile the published imaging demonstrations of thermal meta-optics. While thermal meta-optical imaging has improved significantly over the years, the image quality is still inferior to that of refractive lenses. The case is similar for visible meta-optics \cite{Huang22}, but there are several promising approaches that may be employed to improve the image quality. 

Computational imaging is one approach to circumvent the limitations of meta-optics, especially with regard to broadband imaging. One method is to design a meta-optic that preserves as much information as possible for all wavelengths. While this optic may not produce a high-quality image directly, the image may be reconstructed using simple computational techniques \cite{Colburn18fullcolor,Bayati21EDOF}. In another approach, an optic designed for single-wavelength illumination (and therefore exhibiting chromatic aberrations) is combined with a computational imaging backend with neural network-based image reconstruction techniques to achieve full-color imaging \cite{Dong24, Liu24}. End-to-end approaches, wherein the physical meta-optic is co-optimized with the computational reconstruction software, have also demonstrated promising results at visible wavelengths \cite{Tseng21,Froch24} but have not yet been adapted for thermal imaging. The combination of meta-optics with machine learning-based image reconstruction techniques for compact, high-quality imaging could readily be adapted for thermal wavelengths, and in fact may be easier to implement due to the monochromatic nature of thermal sensors and their limited resolution.  

Imaging applications often desire large aperture, wide FoV, and broadband operation, but we note that achieving the three simultaneously via a single layer of meta-optics has not been demonstrated at any wavelength. Stacking multiple meta-optics in series has been used to correct both monochromatic and chromatic aberrations \cite{Arba16,Groever17,Park23,Shrestha23,Pan23}, but at the cost of increasing the thickness of the optical system. Hybrid systems consisting of a refractive lens with a meta-optic have been experimentally demonstrated at visible wavelengths \cite{Pinilla23} and also recently for thermal wavelengths \cite{Shih22,Liu24hybrid} as a way to achieve broadband imaging with wider FoV. Such combinations of multiple elements may be used to achieve higher quality imaging without the aid of computational reconstruction in applications where such reconstruction is not feasible.

While meta-optical imaging quality is rapidly improving, we note that thermal imaging applications often prioritize functionality over aesthetic imaging, and therefore meta-optics are well-suited to the needs of thermal imaging. In particular, the ability to image at different magnification factors via polarization selectivity \cite{Saragadam24,Ou2021broadband} and axial translation \cite{WirthSingh2024zoom} is especially applicable for long-range sensing applications, where wide FoV is required for context and only a small area must be imaged at high resolution. Large aperture is an another frequently desired aspect of thermal imaging systems in order to increase the signal-to-noise ratio over long range or low-signal environments. For large-area meta-optics, the weight increases negligibly with increased aperture, whereas with traditional refractive optics large aperture contributes significant weight due to the bulk material. While this review has focused primarily on imaging, we note that spectral sensing is another application area of meta-optics that may be especially applicable to thermal wavelengths. By engineering meta-optics to have a particular spectral response, the chromaticity of meta-optics can be leveraged to perform spectral sensing \cite{Froech22,Faraji18,Zhu17}, which has been demonstrated in the visible and near-infrared but not yet at thermal wavelengths.

In conclusion, we have reviewed approaches to thermal imaging via meta-optics, including standard hyperboloid metalenses, wide FoV meta-optics, and broadband meta-optics. With the relatively high cost of traditional optical components and the various functionalities that are enabled by meta-optics, meta-optics are an attractive alternative to traditional components in thermal imaging systems. 

\section*{Funding}
Funding for this work was supported by the federal STTR program and DARPA (HR001123C0034).


\section*{Data Availability Statement}

The data that support the findings of this study are available from the corresponding author upon reasonable request.





\printbibliography[title={References}]  

\end{document}